# X-ray and molecular dynamics study of the temperature-dependent structure of molten NaF-ZrF$_4$

Anubhav Wadehra[1,#], Rajni Chahal[2], Shubhojit Banerjee[3], Alexander Levy[4], Yifan Zhang[5], Haoxuan Yan[1], Daniel Olds[6], Yu Zhong[5], Uday Pal[1,4], Stephen Lam[3], Karl Ludwig[1,7,#]

[1] Division of Materials Science and Engineering, Boston University, 15 St. Mary's St., Boston, Massachusetts 02215, USA

[2] Chemical Science Division, Oak Ridge National Laboratory, Oak Ridge, Tennessee 37831, USA

[3] Department of Chemical Engineering, University of Massachusetts Lowell, Lowell, Massachusetts 01854, USA

[4] Department of Mechanical Engineering, Boston University, 15 St. Mary's St., Boston, Massachusetts 02215, USA

[5] Department of Mechanical and Materials Engineering, Worcester Polytechnic Institute, 100 Institute Rd., Worcester, Massachusetts 01609, USA

[6] National Synchrotron Light Source II, Brookhaven National Laboratory, Upton, New York 11793, USA

[7] Department of Physics, Boston University, 590 Commonwealth Ave., Boston, Massachusetts 02215, USA

[#] Corresponding authors:

  Karl Ludwig - ludwig@bu.edu

  Anubhav Wadehra - awade9@bu.edu

## Abstract

The local atomic structure of NaF-ZrF$_4$ (53-47 mol%) molten system and its evolution with temperature are examined with x-ray scattering measurements and compared with *ab-initio* and Neural Network-based molecular dynamics (NNMD) simulations in the temperature range 515-700 °C. The machine-learning enhanced NNMD calculations offer improved efficiency while maintaining accuracy at higher distances compared to *ab-initio* calculations. Looking at the evolution of the Pair Distribution Function with increasing temperature, a fundamental change in the liquid structure within the selected temperature range, accompanied by a slight decrease in overall correlation is revealed. NNMD calculations indicate the co-existence of three different fluorozirconate complexes: [ZrF$_6$]$^{2-}$, [ZrF$_7$]$^{3-}$, and [ZrF$_8$]$^{4-}$, with a temperature-dependent shift in the dominant coordination state towards a 6-coordinated Zr ion at 700°C. The study also highlights the metastability of different coordination structures, with frequent interconversions between 6 and 7 coordinate states for the fluorozirconate complex from 525°C to 700°C. Analysis of the Zr-F-Zr angular distribution function reveals the presence of both "edge-sharing" and "corner-sharing" fluorozirconate complexes with specific bond angles and distances in accord with previous studies, while the next-nearest neighbor cation-cation correlations demonstrate a clear preference for unlike cations as nearest-neighbor pairs, emphasizing non-random arrangement. These findings contribute to a comprehensive understanding of the complex local structure of the molten salt, providing insights into temperature-dependent preferences and correlations within the molten system.

## Introduction

After the initial studies on molten salt properties performed at Oak Ridge National Laboratory (ORNL) for the Aircraft Reactor Experiment (ARE) and Molten Salt Reactor Experiment (MSRE) starting in the 1950s, and the subsequent development of the Molten Salt Breeder Reactor (MSBR) design later in the 1960s, interest in molten salt technologies weakened[1,2]. The desirable properties of molten salts such as high volumetric heat capacity, low vapor pressure, high boiling point, ability to dissolve actinides and insensitivity to radiation has, however, generated a renewed interest in using them not only in MSRs as fuel carriers, but in Concentrated Solar power (CSP), thermal storage systems (TES), hydrogen production, oil refineries and as heat transfer fluids (HTF) [1,3,4,5,6,7,8]. Since the establishment of Generation IV International Forum in early 2000s, an area of high interest has been to further advance Molten Salt Reactor (MSR) technology since it provides unique opportunities to create a safer, economical, and sustainable method of energy generation as well as supply high temperature industrial heat at low pressures [8,9]. The ability of molten salts to be used as both coolant and fuel carrier and to allow for online reprocessing of the salt in MSRs offer significant advantages [10,11].

While 'molten salts' is a blanket term consisting of, but not limited to, halides, oxides, nitrates and carbonates, fluorides are commonly proposed for thermal-spectrum MSR technologies owing to their combination of neutronic, thermal, and chemical properties. Their lower vapor pressure over a wide range of operating temperatures, relatively large volumetric heat capacities (comparable to water), relatively low melting points and high boiling points make them an excellent candidate for secondary coolant while their better neutron economy, moderating efficiency, chemical stability at higher temperatures, and compatibility with graphite along with low cost of processing makes them an almost unrivaled primary coolant and/or fuel salt carrier[12,13,14,15].

One such salt, NaF-ZrF$_4$, a mixture that was used as the fluoride solvent in the ARE with UF$_4$ as fuel (NaF- ZrF$_4$-UF4 (53-41-6 mol %)), has garnered significant interest as a low-cost baseline salt due to its high solubility for actinides, ability to work as an oxygen getter and good thermal properties [8,16,17]. Additionally, its versatility makes it suitable to be used as a fuel carrier, a primary coolant and/or a secondary coolant. However, a comprehensive understanding of its structural arrangement at elevated temperatures has remained a scientific challenge due to the hazard of working with radioactive fuel along with the presence of various transuranic elements and fission products in the salt mix at any moment. To establish a model system, it is essential to understand the structure and speciation of the salt by itself as properties such as viscosity, solubility and chemical reactivity are influenced by its molecular structure [8]. While there exists a range of compositions of the NaF-ZrF$_4$ system that could be considered, here we have chosen to study 53-47 mol% given its low melting point and closeness to the composition used in ARE without the radioactive element. The creation of advanced x-ray scattering techniques over 5 decades since the ORNL's MSRE, has allowed us to investigate the structure of the salt from 515 °C - 700 °C. Using the X-ray scattering spectra and pair distribution function, closely interfaced with *Ab-Initio* Molecular Dynamics (AIMD) and Neural Network-based Molecular Dynamics

(NNMD), we provide a deeper understanding of the structural motifs in the liquid - beyond coordination numbers and distances.

## Experimental Details

A schematic of the custom x-ray chamber along with the sample holder assembly used to perform the experiments is shown in Fig.1. The sample holder assembly consists of a 3D printed ceramic holder with a through hole for incoming and scattered x-rays. Positioned just above the incoming x-ray window is a notch designed to insert a thermocouple probe (K-type, Omega Engineering) for precise temperature measurement. A Kanthal heating wire (28mm gauge) is wrapped and fastened around the ceramic holder using alumina paste to provide thermal and electrical insulation along with uniform heating to the sample. A cylindrical glassy carbon crucible (HTW Germany) with dimensions of 5mm OD, 8mm height and 0.5 mm wall thickness was used to contain the salt. The crucible, being chemically inert towards the salt and sufficiently transparent to x-rays, was placed within the ceramic holder in the center of the x-ray windows. The holder assembly was mounted to the chamber using a 4.5" CF flange. Following evacuation to 1E-3 psi (50 mtorr), the chamber was backfilled with Ultra High Purity Argon gas at +1 psi relative to ambient to obtain an inert atmosphere. Kapton windows were used for x-ray entrance and exit from the chamber.

NaF-ZrF$_4$ (53-47 mol%) salt was prepared by repeated melting and cooling of the component salts (Sigma Aldrich 99.99% purity NaF, 99.9% purity ZrF$_4$) in a Nickel crucible inside an ultra-high purity Nitrogen glovebox with oxygen and moisture traps to minimize the formation of oxyfluorides. The salt was then crushed to obtain powder samples which were then added to the crucibles, melted, and solidified inside the glove box. The final composition of the salt was confirmed using XRD, SEM-EDS and ICP-OES.

X-ray scattering experiments were performed at beamline 28-ID-1 of the National Synchrotron Light Source-II at Brookhaven National Laboratory using an x-ray energy of 74.5 keV. A Perkin-Elmer flat panel amorphous Silicon area detector with 200 x 200 µm square pixels measured the scattered x-rays at a position 589 mm behind the sample. The scattering geometry was calibrated using a cerium dioxide powder sample inside a glassy carbon crucible. Scattering from NaF-ZrF$_4$ was measured every 25°C from 525 °C to 700 °C after an initial scan at 515 °C. Background scattering was measured for an empty glassy carbon crucible and Kapton for subtraction from the total observed scattering.

## Experimental Results

The program PyFAI was utilized to azimuthally integrate the two-dimensional (2D) x-ray scattering data to one-dimensional (1D) diffraction patterns over an appropriately masked area of the detector [18]. Background scattering was subtracted from the total scattering and the resulting

NaF-ZrF$_4$ scattering patterns were then analyzed using PDFGetX2 with $q_{min}$ = 0.32 Å$^{-1}$ and $q_{max}$ = 13.5 Å$^{-1}$ to preserve information while reducing noise and unphysical features [19]. They were normalized to a per atom basis to obtain the Total Scattering Structure Functions *S(q)* and Reduced Structure functions *F(q)*:

$$F(q) \equiv q[S(q) - 1] = q\left[\frac{I(q)}{\langle f(q) \rangle^2} - \frac{\langle f^2(q) \rangle - \langle f(q) \rangle^2}{\langle f(q) \rangle^2} - 1\right] \quad (1)$$

Here the averages are over the atomic species α, of concentrations c$_α$, and the f$_α$(q) are the atomic scattering factors for the given atomic species. The second term on the right-hand-side is the Laue diffuse scattering. Experimental results extrapolated to *q*=0 and multiplied by a Lorch function sin(Δ$_0$q)/(Δ$_0$q) with Δ$_0$=π/q$_{max}$ to minimize truncation oscillations in subsequent Fourier sine transforms are shown in Fig. 2 [20]. A double peak at low wavenumbers (1-2 Å$^{-1}$) suggests the presence of intermediate range order in the system, likely due to the presence of complexing around the Zr in the system. It can be noted that there are some unusual small features in the experimental F(q) curves near 7.5 Å$^{-1}$. These seem to be related to imperfect background subtraction and do not affect the final quality of the results.

The Reduced Structure Functions F(q) are related to the real-space distribution functions by:

$$F(q) = \sum_{\alpha,\beta} \frac{c_\alpha c_\beta f_\alpha(q) f_\beta^*(q)}{\langle f(q) \rangle^2} \rho_0 \int 4\pi r \left[g_{\alpha\beta}(r) - 1\right] \sin(qr) \, dr \quad (2)$$

where $c_\alpha$ and $c_\beta$ refer to atomic fraction of element species $\alpha$ and $\beta$, $\rho_0$ is the atomic density, $q$ and $g_{\alpha\beta}(r)$ are the scattering wavenumber and the Radial Distribution Functions (RDFs) of $\alpha$-$\beta$ ion pair, respectively; $f_\alpha(q)$ and $f_\beta^*(q)$ are the atomic x-ray scattering factors; $\langle f(q) \rangle$ is the average scattering factor and calculated by:

$$\langle f(q) \rangle = \sum_\alpha c_\alpha f_\alpha(q) \quad (3)$$

The species-specific atomic pair distribution functions $g_{\alpha\beta}(r)$ are:

$$g_{\alpha\beta}(r) \equiv \frac{1}{4\pi c_\alpha c_\beta N \rho_0 r^2} \sum_{\substack{i \in \{\alpha\} \\ j \in \{\beta\} \\ i \neq j}} \delta(r - r_{ij}) \quad (4)$$

where N is the total number of atoms. Because of the concentration- and scattering factor-dependence shown in Eq. 2, the strongest contributions to the x-ray scattering are from Zr-F and F-F and Zr-Zr correlations while Na-F, Na-Zr and Na-Na contributions are quite small.

The experimental Pair Distribution Function (PDF) is the sine transform of F(q):

$$G(r) = \frac{2}{\pi} \int_0^\infty F(q) \sin qr \, dq \quad (5)$$

The experimental PDF as a function of temperature from 515 °C – 700 °C are shown in Fig. 3 and exhibit a consistent trend as temperature is increased. Atomic densities are calculated from the correlations presented by Cohen et al [21]. The G(r) function removes the average density and is particularly useful for examining structure on length scales beyond nearest neighbor. However, for visualization of structure at short distances, it's often useful to examine the Total Correlation Function:

$$T(r) = 4\pi r \rho_0 + \frac{2}{\pi} \int_0^\infty F(q) \sin qr \, dq \qquad (6)$$

As an instructive comparison, Fig. 4 shows the experimental and NNMD T(r) functions at 700 °C at low-r along with the individual pair contributions obtained through NNMD. The shortest correlations in the melt are from Zr-F, Na-F and F-F nearest neighbor distances. As can be seen, the asymmetry as well as overlap between different atomic pair contributions make it difficult to assign a peak in T(r) to a specific atomic pair. Even the strong contribution due to Zr-F correlation in the first peak at ~2 Å is not the only contribution; it also contains Na-F as well as F-F pair contributions. The second prominent peak at ~4.2 Å has a major contribution from the Zr-F second shell along with overlapping contributions from all other atomic pairs, with negligible contribution from Na-Na correlation.

In Fig. 5, we show a purely experimental fit generated using NXfit [22]. The program assumes gaussian coordination shells, sums the partial correlations calculated in Q space and Fourier transforms to real space for comparison with the experimental data. It then provides optimized coordination numbers and distances and disorder parameters for each correlation. Notably, the assumption of symmetrical gaussian contributions results in fits that provide higher coordination distances than those predicted by MD, in which the asymmetrical contributions continue for longer distances (Table 1). Moreover, in experimental fitting, the F-F and Na-F shells emerge after ~2 Å leading to the Zr-F correlation nearly coinciding with the first peak of T(r) indicative of its exclusive contribution to T(r). However, upon comparison with MD T(r), we observe that the rise of F-F and Na-F shells precedes 2 Å, suggesting their small but non-trivial contribution to the first peak of T(r). This subsequently causes a slight deviation in the T(r) peak position compared to the Zr-F peak. As a result, we conclude that the individual contributions to the experimental T(r) function can be fit, but the asymmetric multiple contributions make this difficult to do in a reliable manner and therefore instead turn to simulation for further guidance.

## MD Simulations

Two different sets of MD simulations were performed for comparison with experimental data. The first set consisted of AIMD simulations. AIMD's accuracy comes at the expense of high computational cost, which makes it infeasible to consider large simulation system size. This particularly can be of concern in liquids where intermediate range ordering exists. To address aforementioned limitations of AIMD with respect to system size, Neural Network Interatomic Potential (NNIP) training on a larger system was performed, as described by Chahal *et al.* [23]. This system was effectively transferred to perform simulations on a larger supercell of NaF-ZrF$_4$ for higher efficiency while maintaining ab initio accuracy. Details are immediately below.

**AIMD**:

AIMD simulations were conducted for a supercell composed of 132 atoms, consisting of 21 Na, 18 Zr, and 93 F. This supercell was chosen to represent a mixture with a NaF-ZrF$_4$ composition of 53-47 mol% at three different temperatures: 525 °C, 600 °C, and 700 °C. The initial configuration of the supercell was generated using the Packmol software [24] with a random structure to ensure that atomic positions did not overlap. Subsequently, the supercell's initial volume, containing the random configuration, was equilibrated using the isothermal-isobaric (NPT) ensemble [25,26,27] in the LAMMPS molecular dynamics package [28]. This equilibration process was carried out for a minimum of 1 nanosecond (ns) to achieve equilibrium volumes and structures at the desired temperatures. The resulting equilibrated structures were then employed as starting points for AIMD simulations under the Born-Oppenheimer approximation, utilizing the Vienna Ab-Initio Simulation Package (VASP) [29,30,31]. For electronic self-consistent calculations, the generalized-gradient-approximation (GGA) with the Perdew-Burke-Ernzerhof (PBE) exchange-correlation functional was adopted [32]. Projector augmented wave (PAW) pseudopotentials provided by VASP for Na ($2p^63s^1$), Zr ($4s^24p^65s^24d^2$), and F ($2s^22p^5$) were utilized [33,34]. A plane wave cutoff of 600 eV with a convergence criterion of $10^{-5}$ eV for electronic self-consistency steps was employed. The AIMD simulations were conducted using a time step of 2 femtoseconds (fs), and a single (gamma) k-point mesh for 10,000 steps. The canonical ensemble (NVT) was maintained using a Nosé-Hoover thermostat [35], and periodic boundary conditions were applied to the system. Dispersion corrections were included through the DFT-D3 method proposed by Grimme [36] to improve agreement with experimental densities. To analyze the simulations, radial distribution functions (RDF) for each ion pair were computed and averaged in Python over the latter half of the simulation steps. These RDFs were then weighted and transformed into normalized X-ray scattering functions and compared with experimental data.

**NNIP-based Molecular Dynamics (NNMD) Simulations**

The trained NNIP as obtained from our previous work on LiF-NaF-ZrF$_4$ molten salt, was implemented in the Large Scale Atomic/Molecular Massively Parallel Simulator (LAMMPS) to study NaF-ZrF$_4$ system [37][23]. The transferability of the NNIP towards NaF-ZrF$_4$ was validated by reproducing the structure of a slightly different salt composition (57 NaF – 43 ZrF$_4$) where previous EXAFS data was available [38]. Additional validation on average coordination of cation species was done using VASP AIMD calculations for this composition. More details on validation of LiF-NaF-ZrF$_4$ NNIP's transferability towards NaF-ZrF$_4$ are provided in the SI. Upon validation, the trained NNIP potential was employed to study 53 NaF-47 ZrF$_4$ mol% composition at 525 °C, 600 °C, and 700 °C. For all temperatures, systems containing 18,632 atoms were simulated at densities obtained from Redlich-Kister (RK) expansion method, which is shown to estimate NaF-ZrF$_4$ salt densities within 1.5% of the experimental values [39]. Here, at corresponding RK densities, the system sizes larger than 65 Å were chosen to allow for accurate representation of fluorozirconate chains that are shown to affect the structure and properties of salt systems containing higher mol% of ZrF$_4$ and was previously found to produce converged structures and transport properties [23]. Simulations were performed with a 1-femtosecond time step for at least 1.5 nanoseconds under the canonical ensemble (NVT) using a Nosé-Hoover thermostat [35] while maintaining the periodic boundary conditions. For all temperatures, equilibrated trajectories of nearly 1 ns were used to perform the structure analysis presented in the next sections.

To compare the MD $g_{\alpha\beta}(r)$ results with experiments, we calculate the predicted x-ray PDF and total correlations functions. For this, Eq. 2 was used to calculate F(q) from g(r), with a Lorch function which truncated the results at $q_{max}$ = 13.5 Å$^{-1}$. Figure 6 compares the results of both MD techniques against the experimental F(q). Similar to experiment, both AIMD and NNMD present the double peak feature found at 1-2 Å$^{-1}$ that points toward intermediate range order, with NNMD more accurately reproducing the peak positions, but AIMD still showing both peaks despite a smaller cell size used for the simulations [40]. To yield the predicted PDF, both the MD F(q) were then Fourier sine transformed back to real space using Eqs. 4 and 5 and compared to experiment as shown in Fig 7. The AIMD predicted PDF agrees well up to ~4.5 Å , while the NNMD predicted PDF agrees very well with experiment for much longer distances overall (10 Å). Figure 8 shows a comparison of individual species-specific pair distribution functions $g_{\alpha\beta}(r)$ obtained from NNMD and AIMD simulations at 525 °C. The dotted line corresponds to AIMD results while the solid line is for NNMD. Data obtained from NNMD matches the AIMD results very well with only a slight difference in the Zr-F correlation. The changes in predicted PDF (obtained from NNMD) as temperature is increased from 525 °C to 700 °C are displayed in Fig. 9 where we see a consistent decrease in correlation as observed in experiment. In sum, we see here that the larger size NNMD reproduces results as good, if not better, than size-limited AIMD (especially at longer distances) for this system which exhibits intermediate-range order; moreover, the larger system size of the NNMD simulations enables better statistical averaging. Consequently, we will focus on NNMD simulations to gain more insight into the structure of the liquid.

## Atomic Structure of the liquid

Nearest-neighbor cation-anion and F-F coordination distances, numbers and root mean square displacements are given in Table 1. The distances are the peaks in the MD cation-anion radial distribution functions $g_{\alpha\beta}(r)$; their uncertainties are approximately ±0.01 Å, based on difficulty of identifying the peak position. The coordination numbers include contributions up to the first minimum in the $g_{\alpha\beta}(r)$ functions.

Figure 8 shows the highest maximum, lowest minimum and smallest width for the Zr-F RDF peak suggesting a strong correlation between Zr and F in the melt while the much broader peak for F-Na suggests comparatively weak correlation among the respective cation–anion pairs. For LiF-NaF-ZrF$_4$ systems, Chahal *et. al* had found a very low minimum in the Zr-F RDF that indicated strong solvation with limited exchange of fluorine from the Zr first solvation shell while a much higher minimum in the Na-F RDF suggested weak solvation allowing a rapid exchange of free and coordinated fluoride ions [40]. The stronger association between Zr and F in the molten salt mixture is also supported by the smaller radius of the first Zr-F peak, $r_{avg}$ = 2.03 Å. A double peak for Zr-Zr correlation is also observed at approximately 3.7 Å and 4.1 Å supporting the presence of significant intermediate ordering effects (see Fig. S1 in SI) that are observed as multiple complexes connected by bridging Fluorine ions that are discussed further in the following paragraphs.

Angular Distribution Functions (ADFs) showing the distribution of bond angles as a function of the polar angle θ provide further information about local cation-anion coordination environment. Changes in [ZrF$_x$] complex distribution in the melt at elevated/different temperatures can be understood from 2D density plots shown in Fig. 10 where $θ_{Zr-F-Zr}$ is plotted against Zr–Zr distances ($d_{Zr-Zr}$). The $θ_{Zr-F-Zr}$ angles are between adjacent fluorozirconate complexes connected by a bridging fluorine atom and $d_{Zr-Zr}$ is the distance between the Zr atoms from the connected fluorozirconates. The histograms along the abscissa and ordinate are the addition of all the bin values along the corresponding directions. The two distinct peaks in $θ_{Zr-F-Zr}$ angle distribution corresponding to the high-density region in the corresponding density plot indicate the presence of edge-sharing and corner-sharing complexes. The number of face-sharing complexes at ~95° is almost negligible at all temperatures as indicated by a sparse region in the density plot and the absence of a distinctive third peak in the angle distribution. The corner-sharing complex average connectivity angle is ~145° with $d_{Zr-Zr}$ of ~4.1 Å while the angle is nearly 115° for edge-sharing complexes with $d_{Zr-Zr}$ ~3.7 Å – similar to observations by Chahal *et al* for LiF-NaF-ZrF$_4$ [40]. The presence of two major complexes was also suggested by the double peak in the Zr-Zr RDF at similar distances. A typical geometry of the 6-, 7- and 8- coordinated zirconium fluoride complexes present in the NNMD predicted melt are displayed in Fig. 11. These typical structures can be considered as disordered octahedral for the 6-coordinated, disordered pentagonal bipyramidal for the 7- coordinated and disordered square antiprism for the 8-coordinated complexes. Examining the Na-F-Na, Na-F-Zr and F-Zr-F bond angle distribution in Fig. 12, we observe a single broad peak in the Na-F-Na system indicating less defined structure between the Na-Na atom pairs while in the Na-F-Zr system, a slight shoulder at ~140° in addition to a single broad peak hint at the presence of an intermediary level of order. As for the F-Zr-F system, distinct peaks emerge at ~80° and a smaller one at ~140° with a slight shoulder at 170°. These correspond very well with the expected bond angles present in disordered octahedral (roughly 90° and 180°) and pentagonal bipyramidal (roughly 72°, 90°, 144° and 180°) structures. A subtle shift in the first peak towards 90° and the second peak towards 180° with increasing temperature can be attributed to the transition from 7- to 6-coordinated complexes.

To quantify the distribution of various coordinate complexes present in the melt, we plot the population of 6 to 8 coordinated [ZrF$_4$] complexes at 525, 600 and 700 °C (from left to right) in Fig. 13. Just above melting temperature, at 525 °C, the dominant contribution is from an almost equal distribution of 6- and 7-coordinated complexes (~28%). However, with increasing temperature, 6-coordinated species become predominant (~32 at 600 °C and 42% at 700 °C) with a decrease in both 7- and 8-coordinated complexes with almost negligible amount of 8-coordinate complex present at 700 °C.

To gain more insight into the presence and stability of the different coordination states in the melt, we explore their relative free energies. Using a smooth function, $f_i$ (o<$f_i$<1), we first obtain the coordination number of $Zr^{4+}$, as shown in Eq. 7:

$$\text{CN} = \sum_{i=1}^{N_F} \frac{1-\left(\frac{r_i}{r_{cut}}\right)^{12}}{1-\left(\frac{r_i}{r_{cut}}\right)^{24}} = \sum_{i=1}^{N_F} f_i \qquad (7)$$

where $r_i$ is the distance between the $i^{th}$ F$^-$ and $Zr^{4+}$ and $r_{cut}$ represents the location of the boundary of the first fluoride coordination shell obtained from the first minimum of the Zr-F $g(r)$ (2.84 Å, in this case) [23]. Equation 7 allows smooth transitions of F$^-$ across the boundary of the first fluoride solvation shell, providing the instantaneous value of CN throughout the MD simulation. $N_F$ is the total number of F$^-$ ions. The instantaneous CNs obtained from the MD trajectories at the temperature T can be used to determine their free energy profile (W(CN)) in terms of their probability distribution (P(n)) via the relation: W(CN) = -$k_B T$ ln[$P(CN)$], plotted as a function of temperature in Fig. 14. Here, the minima and barriers between them quantify the multiplicity and metastability of CN. The NNMD simulations suggest a preference for the 7-coordinate state at 525 °C, characterized by its lowest free energy, followed by the 6-coordinate state. At this temperature, the system is required to overcome a barrier of 3.5 kcal/mol to transit from the 7-coordinate state to the 6-coordinate state compared to 4 kcal/mol for the 8-coordinate state. As temperature is increased to 600 °C, a shift is observed, with the system exhibiting a preference for the 6-coordinate state with a similar barrier (3.5 Kcal/mol) required to transition to 7-coordinate state. This barrier increases to ~4 kcal/mol at 700 °C. For all temperatures, the lower barrier values between 6- and 7-coordinate state facilitate frequent interconversions while the larger barrier values between 7- and 8-coordinate states limit the occurrence of 8-coordinate complexes.

Turning now to the next nearest neighbor (NNN) cation-cation correlations, Table 2 shows the calculated number of cation-cation NNNs at 525 °C out to the first minimum of $g_{\alpha\beta}(r)$. The right column shows the cation fractions of NNN for each cation central atom; these can be compared with the NaF-ZrF$_4$ Na-Zr cation fraction of 0.53-0.47. Relative to the random composition values, there is a clear preference for the unlike cation to be the NNNs.

## Conclusion

Through a combination of high temperature experimental and NNMD-calculated PDF, the local structure in the prototypical NaF-ZrF$_4$ molten system was investigated for the first time, with high spatial resolution. Careful consideration of the evolution of the pair distribution functions with increasing temperature calculated through MD show fundamental change in liquid structure in the selected temperature range along with slight decrease in overall correlation. In accord with previous studies by Pauvert et. al. (LiF-ZrF$_4$ mixtures) and Chahal et. al. (LiF-NaF-ZrF$_4$ mixtures), the NNMD calculations presented here indicate the co-existence of three different fluorozirconate complexes: $[ZrF_6]^{2-}$, $[ZrF_7]^{3-}$ and $[ZrF_8]^{4-}$ in the NaF-ZrF$_4$ salt mix [40,41]. A temperature-dependent shift in the dominant coordination state is revealed with a decrease in the average Zr coordination number with increase in temperature leading to a 6-coordinated Zr ion dominating at 700 °C. Moreover, a low free energy barrier value highlights the metastability of different coordination structures with frequent interconversions between the 6 and 7 coordinate states for the fluorozirconate complex from 525 °C to 700 °C.

Further insight into the local structure is gained by analyzing the Zr-F-Zr angular distribution function. The presence of both edge-sharing and corner-sharing fluorozirconate complexes in approximately equal quantity with specific bond angles and distances is evident with a negligible number of face-sharing complexes present. Additionally, the study explores the NNN cation-cation correlations revealing a clear preference for unlike cations to be the nearest-neighbor cation-cation pairs emphasizing the non-random composition of the pairs.

Overall, the findings contribute to a comprehensive understanding of the complex local structure of the molten salt, shedding light on temperature-dependent preferences and correlations in the molten system.

| | NXFit R | NXFit CN | NNMD R | NNMD CN |
|---|---|---|---|---|
| Zr-F | 2.04 | 7.00 | 2.00 | 6.5 |
| Na-F | 2.53 | 6.98 | 2.30 | 7.8 |
| F-F | 2.98 | 7.54 | 2.75 | 9.5 |

Table 1. NN coordination distances and numbers at 525 °C

| Central Atom \ Coordinating Atom | Na | Zr | Total NNN CN | Na-Zr Fractions 53-47 |
|---|---|---|---|---|
| Na | 4.7 | 6.3 | 11 | 0.43 – 0.57 |
| Zr | 7.1 | 2.7 | 9.8 | 0.72 – 0.28 |

Table 2. NNN cation-cation coordination numbers and atomic species fractions at 525 °C

## Acknowledgements

This research at Boston University, University of Massachusetts-Lowell and the Worcester Polytechnic Institute was supported by the US Department of Energy (DOE) NEUP program [Grant Number 20-19373] and Award DE-NE0009204, and the US National Science Foundation (NSF) [Award No. CMMI-1937818 and CMMI 1937829]. This research used resources of the National Synchrotron Light Source II, a U.S. Department of Energy (DOE) Office of Science user facility operated for the DOE Office of Science by Brookhaven National Laboratory under Contract No. DE-SC0012704.

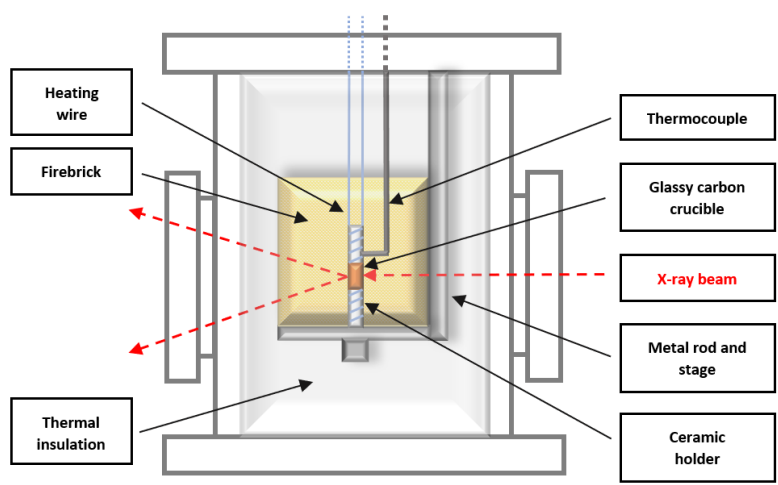

*Figure 1.* Schematic of sample cell and x-ray chamber used to house the molten salt during experiments at beamline 28-ID-1, NSLS II, Brookhaven National Laboratory.

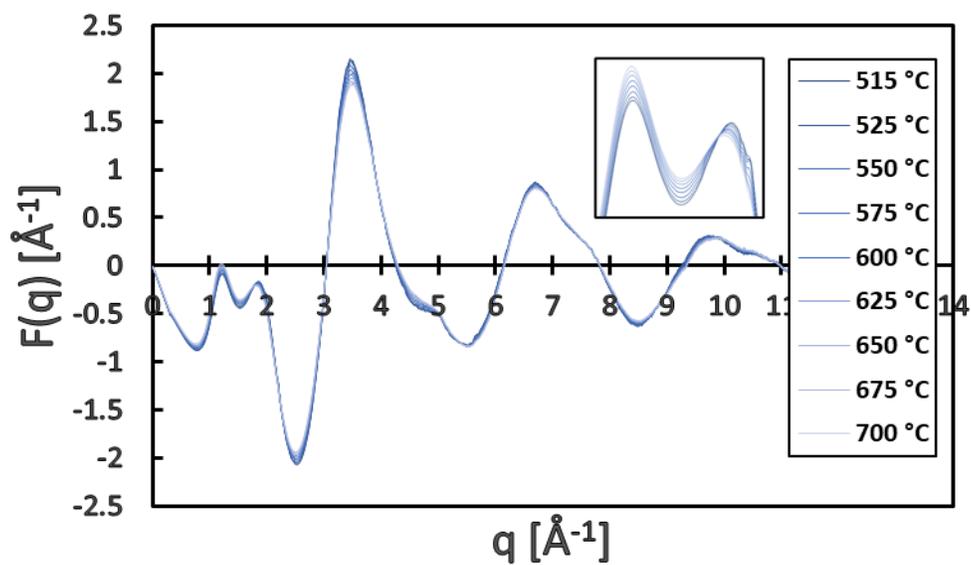

*Figure 2.* Evolution of x-ray Reduced Structure Function F(q) with temperature.

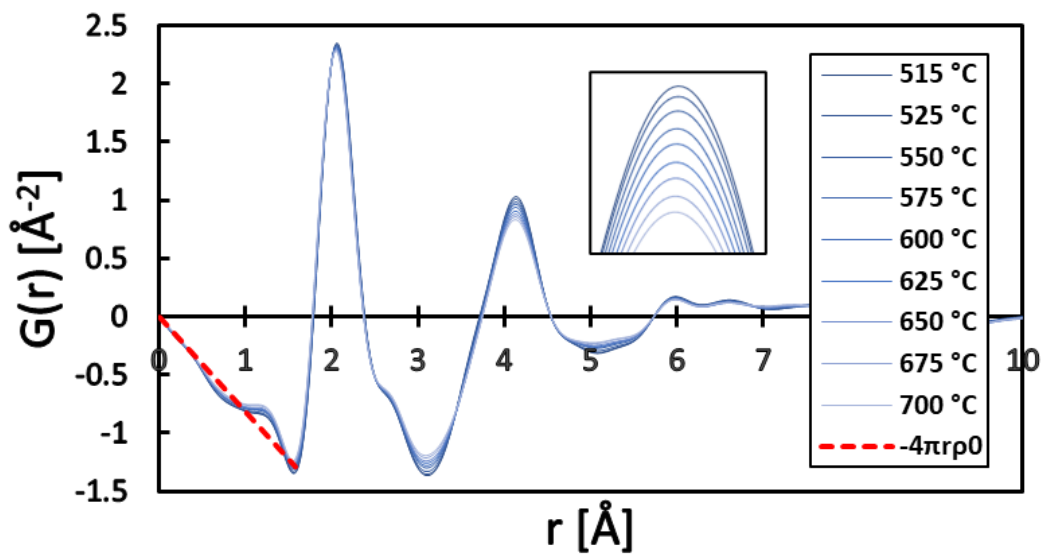

**Figure 3.** Evolution of experimental Pair Distribution Function G(r) with temperature.

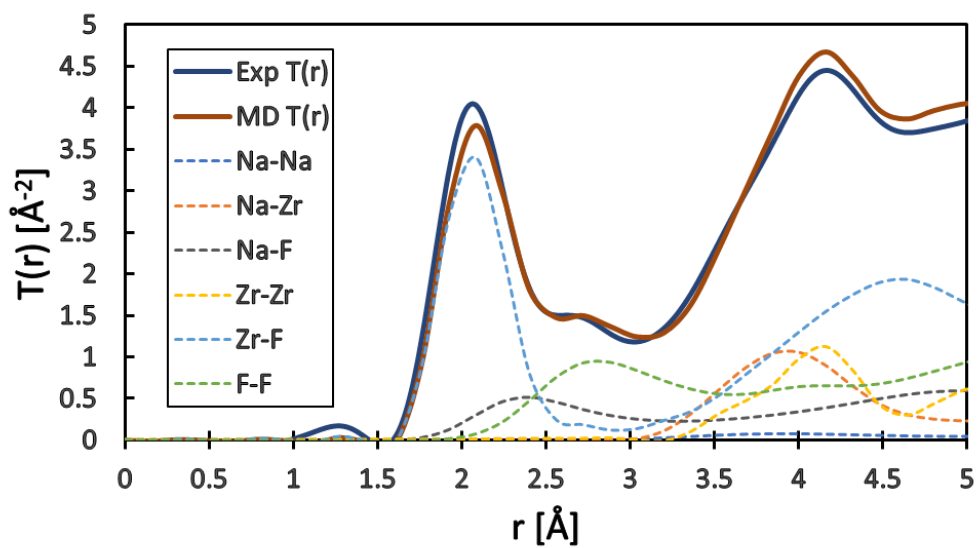

**Figure 4.** Experimental Total Correlation Function T(r) at 525 °C and comparison with T(r) calculated from NNMD simulations using the same structure factor weightings and transform cutoff as for the experiment. Also shown from MD are contributions of prominent species pair correlations to T(r).

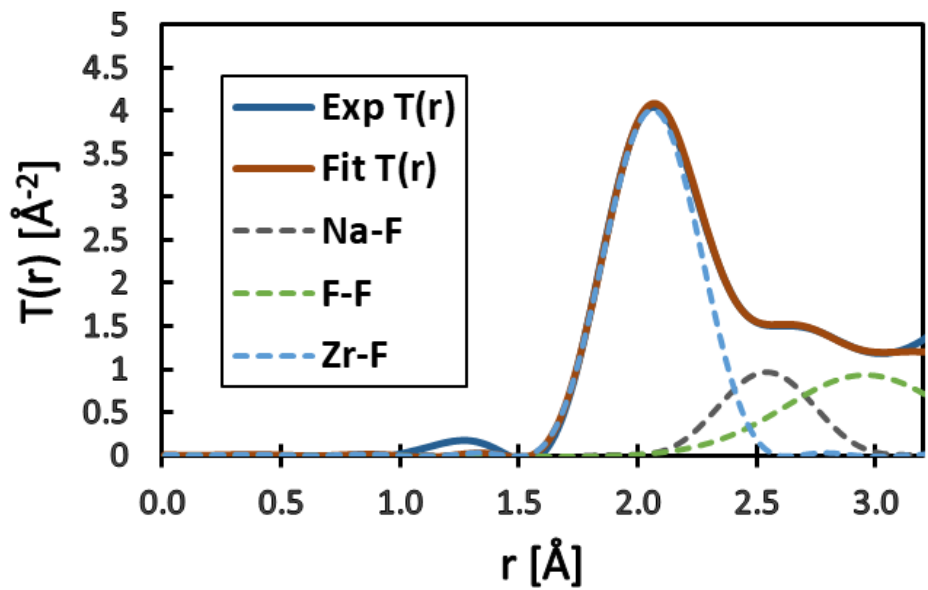

**Figure 5.** Experimental Total Correlation Function T(r) at 525 °C fit using NXFit to obtain structural parameters. Also shown are contributions of prominent species pair correlations to the T(r) fit up to 3.2 Å.

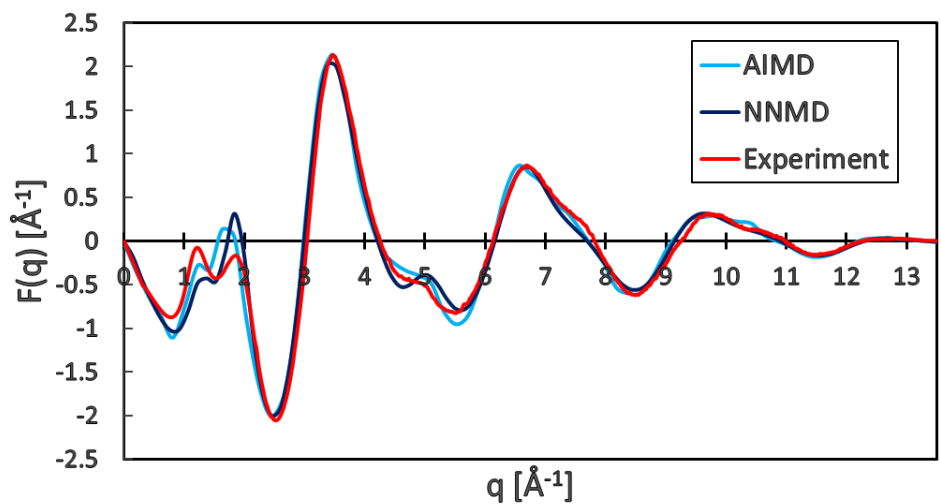

**Figure 6.** Experimental F(q) and comparison with Predicted F(q) calculated from AIMD and NNMD at 525 °C.

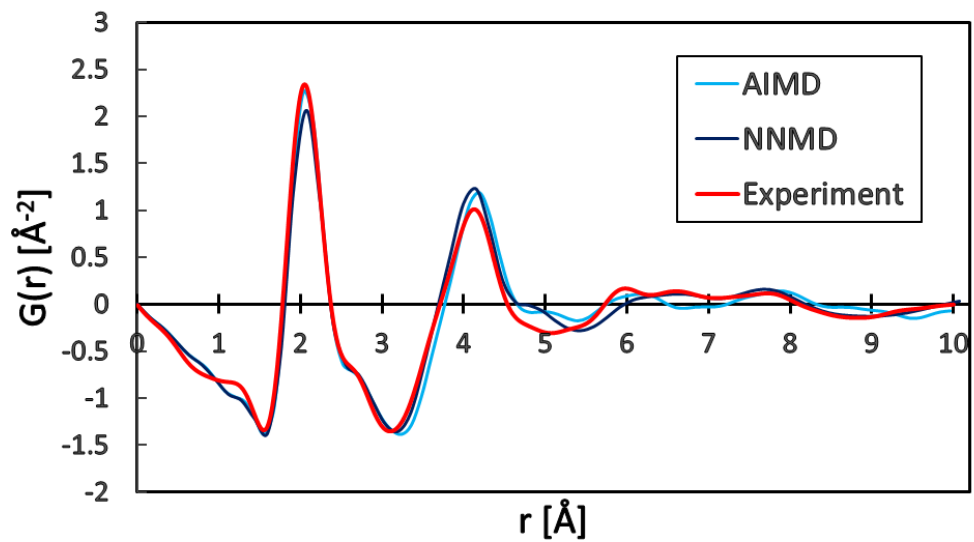

*Figure 7.* Experimental G(r) and comparison with Predicted G(r) calculated from AIMD and NNMD species-specific pair distribution functions at 525 °C.

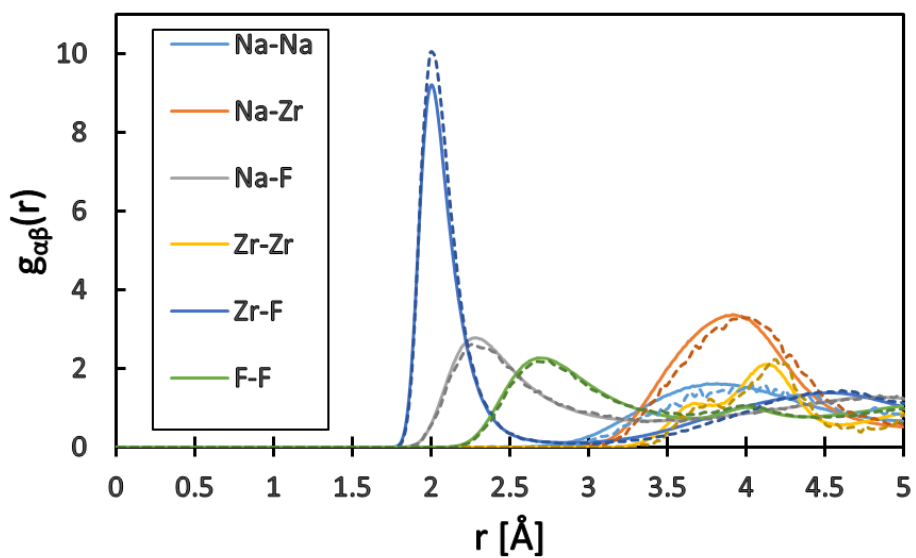

*Figure 8.* Comparison of individual species-specific pair distribution functions between NNMD and AIMD simulations at 525 °C. The dotted line shows AIMD while the solid line is for NNMD.

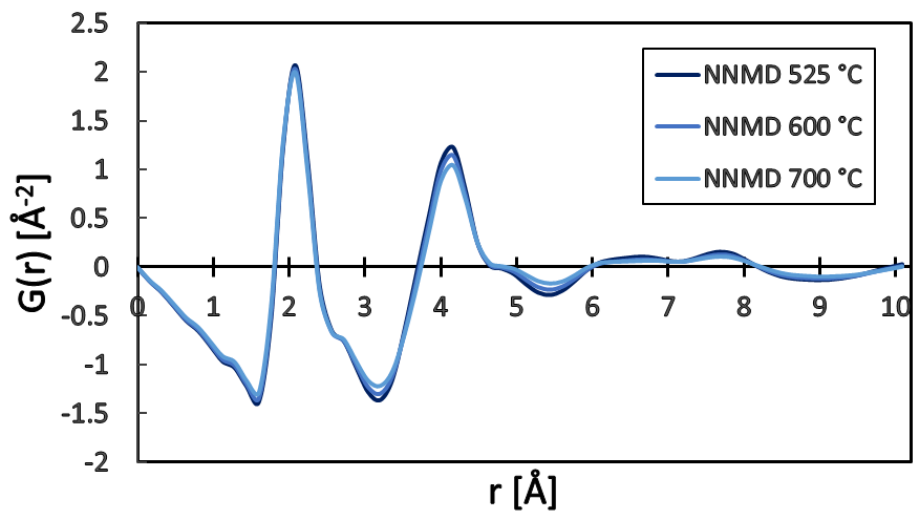

*Figure 9.* Predicted x-ray Pair Distribution Function G(r) calculated from the NNMD species-specific pair distribution functions as a function of temperature.

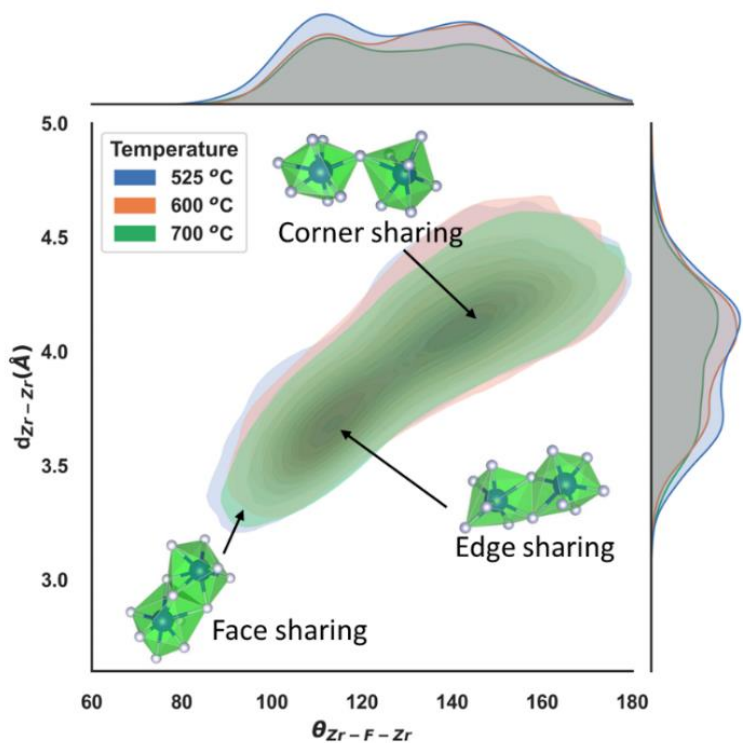

*Figure 10.* Zr-F-Zr bond angle distribution from NNMD calculations. The number of corner sharing complexes is higher at low-temperature.

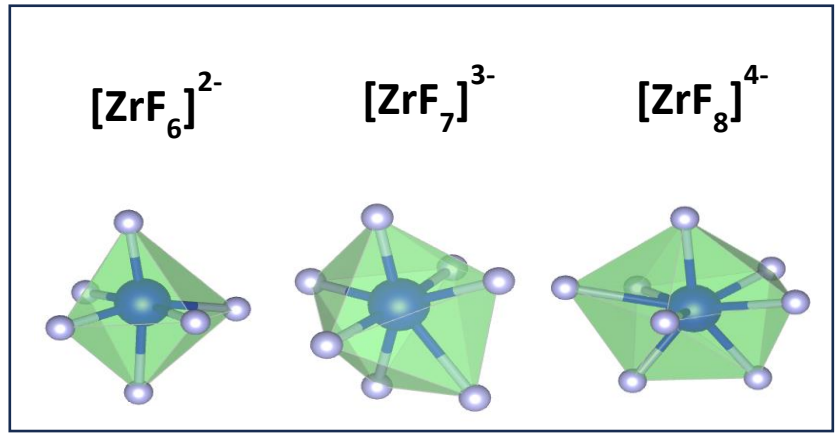

*Figure 11. Snapshot of a typical geometry of the 6-, 7-, and 8-coordinated zirconium fluoride complexes present in the NNMD predicted melt.*

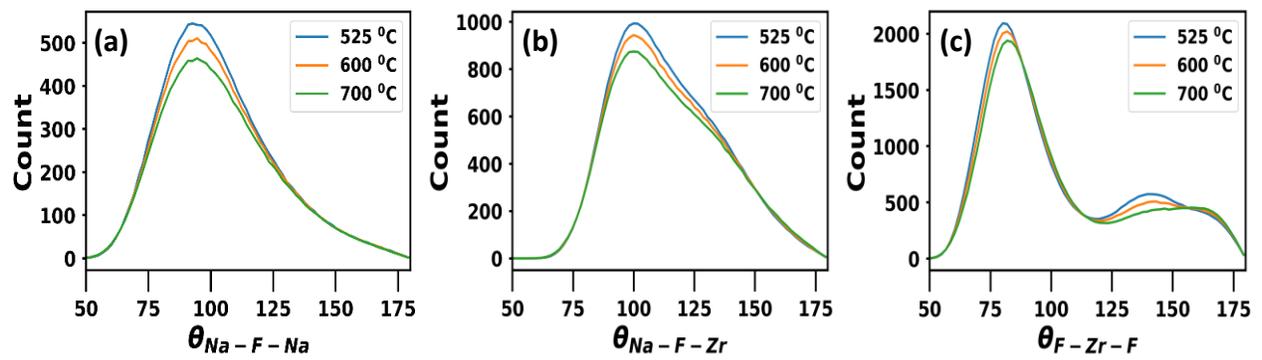

*Figure 12. (a) Na-F-Na, (b) Na-F-Zr and (c) F-Zr-F bond angle distribution from NNMD calculations.*

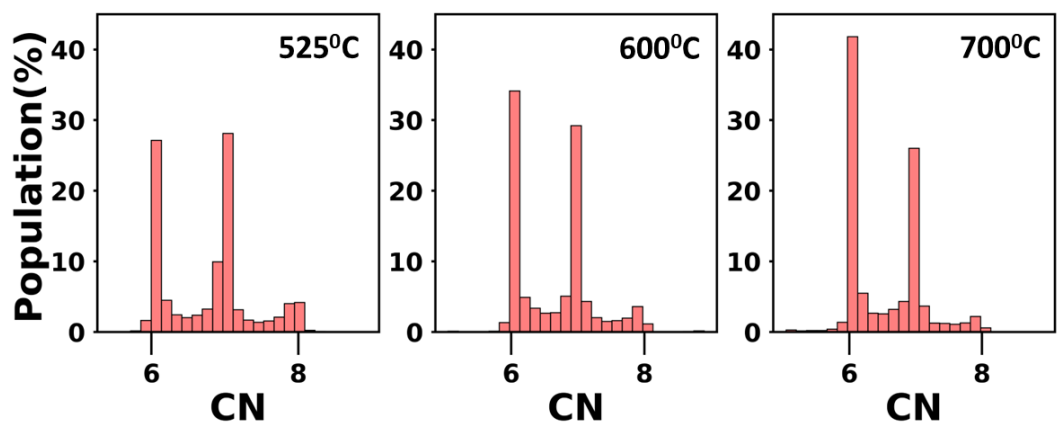

*Figure 13.* Population distribution of Zr-F coordination numbers (CN) at 525 °C, 600 °C and 700 °C

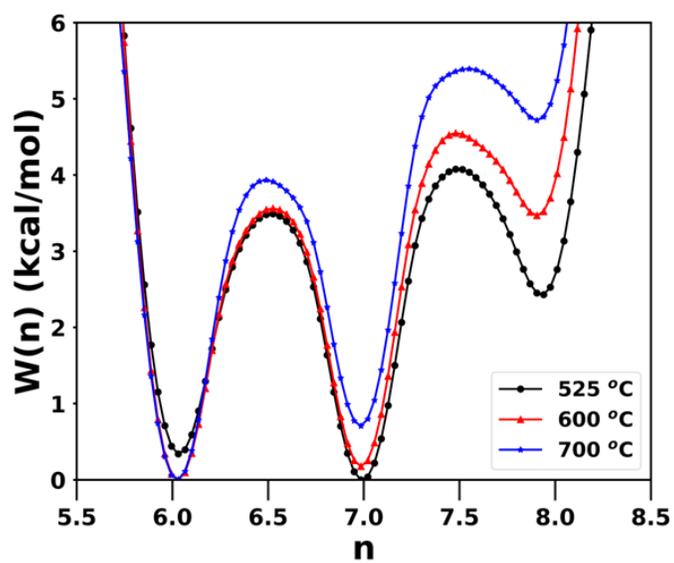

*Figure 14.* Free energy profiles highlighting the distributions and stability of 6-, 7- and 8- fluoride coordination number of $Zr^{4+}$ for different temperatures obtained from NNMD simulations.

# Supporting Information

# X-ray and molecular dynamics study of the temperature-dependent structure of molten NaF-ZrF$_4$

**Validating LiF-NaF-ZrF$_4$ NNIP for NaF-ZrF$_4$ System**

In this work, the trained NNIP for LiF-NaF-ZrF$_4$ molten salt in our previous study [1] was implemented in the Large Scale Atomic/Molecular Massively Parallel Simulator (LAMMPS) to study 53 NaF-47 ZrF$_4$ system. For this purpose, the transferability of the potential energy surface (PES) generated for LiF-NaF-ZrF$_4$ needs to be determined. As experimental EXAFS data existed for 57–43 mole% NaF-ZrF$_4$ salt composition [2], the developed NNIP was employed to study the coordination complexes and average coordination numbers of cation species for this composition. To evaluate average coordination numbers for NNIP's validation, a short VASP AIMD equilibrated trajectory was also used. For VASP calculation at 973 K, 11.35 Å simulation cell containing 16 Na, 12 Zr, and 64 F atoms was considered. Born–Oppenheimer AIMD simulations were performed with the projector-augmented wave (PAW) method, a plane-wave basis set, and the Perdew–Burke–Ernzerhof (PBE) generalized-gradient-approximation (GGA) exchange correlation functional [3,4]. PAW-PBE potentials provided by VASP were used for Na_sv ($2s^2 2p^6 3s^1$), Zr_sv ($4s^2 4p^6 4d^2 5s^2$), and F_s ($2s^2 2p^5$). A plane-wave basis set with an energy cutoff of 650 eV was used. The convergence criterion of 1E-5 eV was set for electronic self-consistent steps. A gamma-centered 1 × 1 × 1 k-point mesh was used for reciprocal space sampling. The parameters chosen yielded convergence within 2 meV/atom and agrees with previous studies [5]. Charges were calculated within the VASP code. The density functional theory (DFT)-D3 formulation proposed by Grimmes [6] was used to account for the effect of dispersion interactions. The canonical ensemble (NVT) using a Nosé–Hoover thermostat [7] was employed. The system was allowed to equilibrate for ~15 ps and afterwards, more than 15 ps of equilibrated trajectory was used to evaluate average coordination numbers for validation of NNIP.

Implementing the existing NNIP for 57 NaF-43 ZrF$_4$ 91 atom system in LAMMPS revealed the existence of 6-, 7-, and 8-coordinated fluorozirconate complexes, which agreed well with both EXAFS measurements and AIMD calculations. Further, as reported in Table S1, the NNIP's prediction of cation coordination with F was also in good agreement with that obtained from AIMD simulations. All calculations were done at experimental density of ~3.042 g/cc at 973 K using 1 fs timestep.

|      | R$_{cut-off}$ | CN, VASP | CN, NNMD |
|------|---------------|----------|----------|
| Zr-F | 2.84          | 6.34     | 6.3      |
| Na-F | 3.48          | 8.36     | 8.3      |

**Table S1.** Validating NNIP for 57 NaF-43 ZrF$_4$ to predict cation coordination.

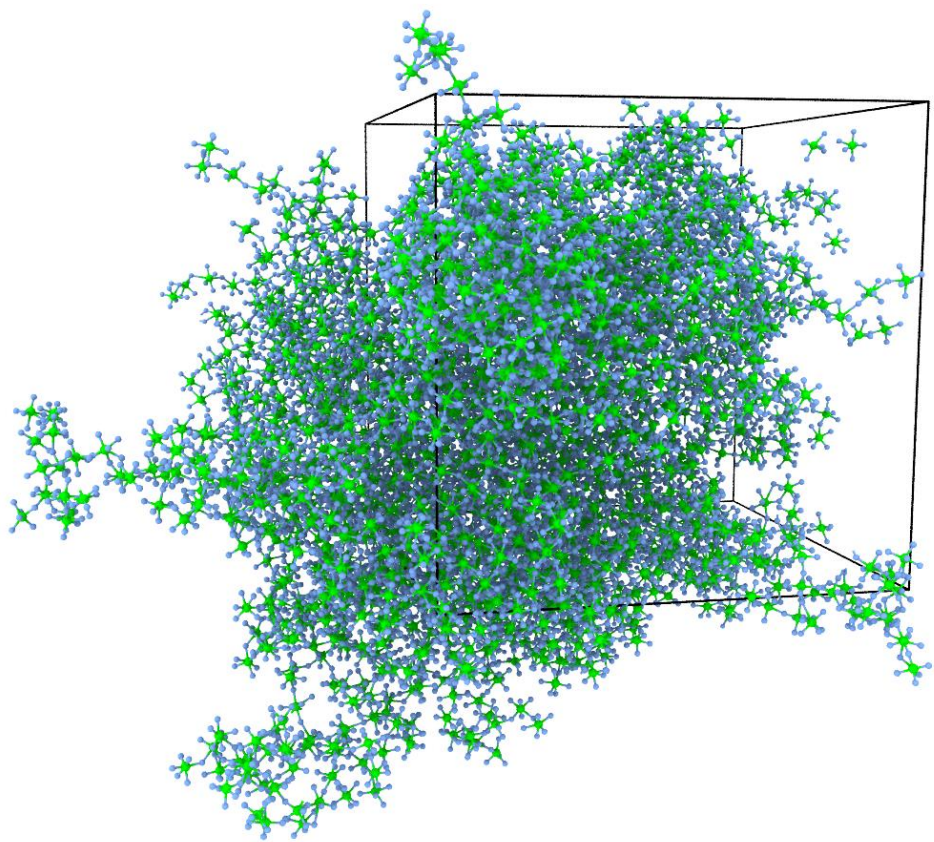

**Figure S1.** Snapshot from NNMD simulations at 973K showing intermediate range ordering in 53 NaF-47 ZrF$_4$ molten salt. Na atoms were removed during visualization in OVITO [8].